\documentclass[11pt, letterpaper]{article}
\usepackage[margin=1in]{geometry}
\usepackage{amsmath,amssymb,amsthm}
\usepackage{graphicx} 
\usepackage{braket}
\usepackage{dsfont}
\usepackage{ulem}
\usepackage[allcolors=blue,colorlinks=true,hyperindex=true]{hyperref}
\usepackage{tabularx}
\usepackage{multirow}
\usepackage{array}
\usepackage{nicematrix}
\usepackage{subcaption}
\usepackage{booktabs}
\usepackage{siunitx}
\title{Quantum Image Classification: Experiments on Utility-Scale Quantum Computers}

\author{Hrant Gharibyan, Hovnatan Karapetyan, Tigran Sedrakyan,\\  Pero Subasic, Vincent P. Su, Rudy H. Tanin, Hayk Tepanyan  \\
\\ {\it BlueQubit Inc, 
San Francisco, CA 94105, USA}
\\ {\it Honda Research Institute USA, 
San Jose, California, 95134, USA}}

\begin{document}

\maketitle
\begin{abstract}
    We perform image classification on the Honda Scenes Dataset on Quantinuum's H-2 and IBM's Heron chips utilizing up to 72 qubits and thousands of two-qubit gates.
    For data loading, we extend the hierarchical learning to the task of approximate amplitude encoding and block amplitude encoding for commercially relevant images up to 2 million pixels. Hierarchical learning enables the training of variational circuits with shallow enough resources to fit within the classification pipeline. For comparison, we also study how classifier performance is affected by using piecewise angle encoding.
    At the end of the VQC, we employ a fully-connected layer between measured qubits and the output classes. Some deployed models are able to achieve above 90\% accuracy even on test images. In comparing with classical models, we find we are able to achieve close to state of the art accuracy with relatively few parameters. These results constitute the largest quantum experiment for image classification to date.
  \end{abstract}
\tableofcontents
\clearpage

\section{Introduction}
Quantum computing is rapidly progressing, with increasing interest in its potential for advantage over classical methods in diverse areas from simulation to quantum chemistry to machine learning. Image classification represents a well established problem in the classical machine learning arena that can serve as a testbed for quantum alternatives. Current generation quantum processors, still within the NISQ era \cite{Preskill2018quantumcomputingin}, are constrained by limited qubit counts, coherence times, and gate fidelities. However, with rapid progress, it is important to benchmark the performance of these devices as we seek to reach the crossover point of quantum utility. 

Supervised machine learning is considered one of the most promising applications for near-term quantum computers~\cite{Dunjko_2018, Biamonte_2017}. Supervised quantum machine learning (QML) algorithms leveraging parameterized quantum circuits (PQCs)~\cite{Cerezo2021-zv, McClean_2016}, such as quantum kernels and variational classifiers \cite{Havlicek2019-zk, lloyd2020quantumembeddingsmachinelearning, Jerbi2023-ft}, have been attracting growing interest in recent years. These methods have been rigorously demonstrated to efficiently tackle specific learning tasks that are intractable for classical approaches~\cite{Liu_2021}. The increasing attention stems from the potential of PQC-based algorithms to offer computational advantages in scenarios where classical methods fall short, particularly in high-dimensional feature spaces and complex problem landscapes. A number of works have found success in hybrid quantum classical architecture~\cite{Senokosov_2024,Liu_2021_hqcnn,Sagingalieva_2023}, though here we primarily focus on quantum classifiers without the classical convoluational layers or feature extraction.

Recent studies have demonstrated the successful implementation of quantum machine learning techniques on actual quantum hardware. In~\cite{johri2021nearest}, the authors developed a nearest neighbor centroid classification algorithm on IonQ's 11-qubit quantum device. In~\cite{Wu_2021}, a variational classifier was applied to LHC event data, achieving performance comparable to classical methods using 10 qubits (with classical data reduced to 10 PCA components). The image classification of the MNIST-Fashion dataset was explored on IBM's 27-qubit device in~\cite{shen2024classificationfashionmnistdatasetquantum}. Additionally, a kernel method was implemented on Google’s Sycamore chip, utilizing 17 qubits, as reported in~\cite{peters2021machine}. A thorough overview of recent experimental results in quantum machine learning methods can be found in~\cite{gujju2024quantum}.

At present, there are a number of hurdles to clear in order to declare a quantum advantage for machine learning. One must show that the learning can be done (e.g., avoiding vanishing gradients, also known as barren plateaus~\cite{McClean_2018-barren}), that the hardware is capable of executing the circuits faithfully, and further that the whole pipeline cannot be classically simulated. We will focus on the first two components. Namely, we will utilize classical simulation on GPU/CPUs to do the training of variational circuits for the purpose of image classification. We then deploy these circuits on quantum hardware—Quantinuum's H1 and H2 chips (with 20 and 56 qubits, respectively) and IBM's Brisbane and Fez systems (with 127 and 156 qubits)—to evaluate the reproducibility of the results. However, by design, the experiments we perform do not constitute something beyond the reach of classical simulation.

A common problem present in machine learning problems in the quantum setting is the question of how to represent the classical data in quantum registers. We explore the performance of a few different loading methods. First, we perform approximate amplitude encoding using a hierarchical circuit ansatz~\cite{gharibyan2023hierarchical,honda1}, which helps mitigate the effect of barren plateaus. We also consider a variant called block amplitude encoding (BAE)~\cite{Nakaji_2022,honda1}, which performs amplitude encoding separately between different parts of the image. Finally, we use a straightforward angle encoding scheme~\cite{Schuld_2019} as well as a method inspired by re-uploading~\cite{PerezSalinas2020datareuploading}. To our knowledge, these are the largest QML experiments done, utilizing qubit numbers up to 72 and thousands of two-qubit gates. 

The structure of the paper is as follows. In Section~\ref{sec:background}, we share details on the image dataset used as well as different methods of encoding the data. We give an overview of image classification performance obtained from simulating quantum circuit classifiers with the different encoding methods and benchmark against classical methods in Section~\ref{sec:simulation}. In Section~\ref{sec:quantum-expts}, we share the results of deploying these circuits on the largest available systems from IBM and Quantinuum. Finally, we conclude in Section~\ref{sec:discussion}. All numerical quantum simulations were orchestrated with the BlueQubit platform.

\section{Background}\label{sec:background}
\begin{figure}
  \centering
  \includegraphics[width=.7\textwidth]{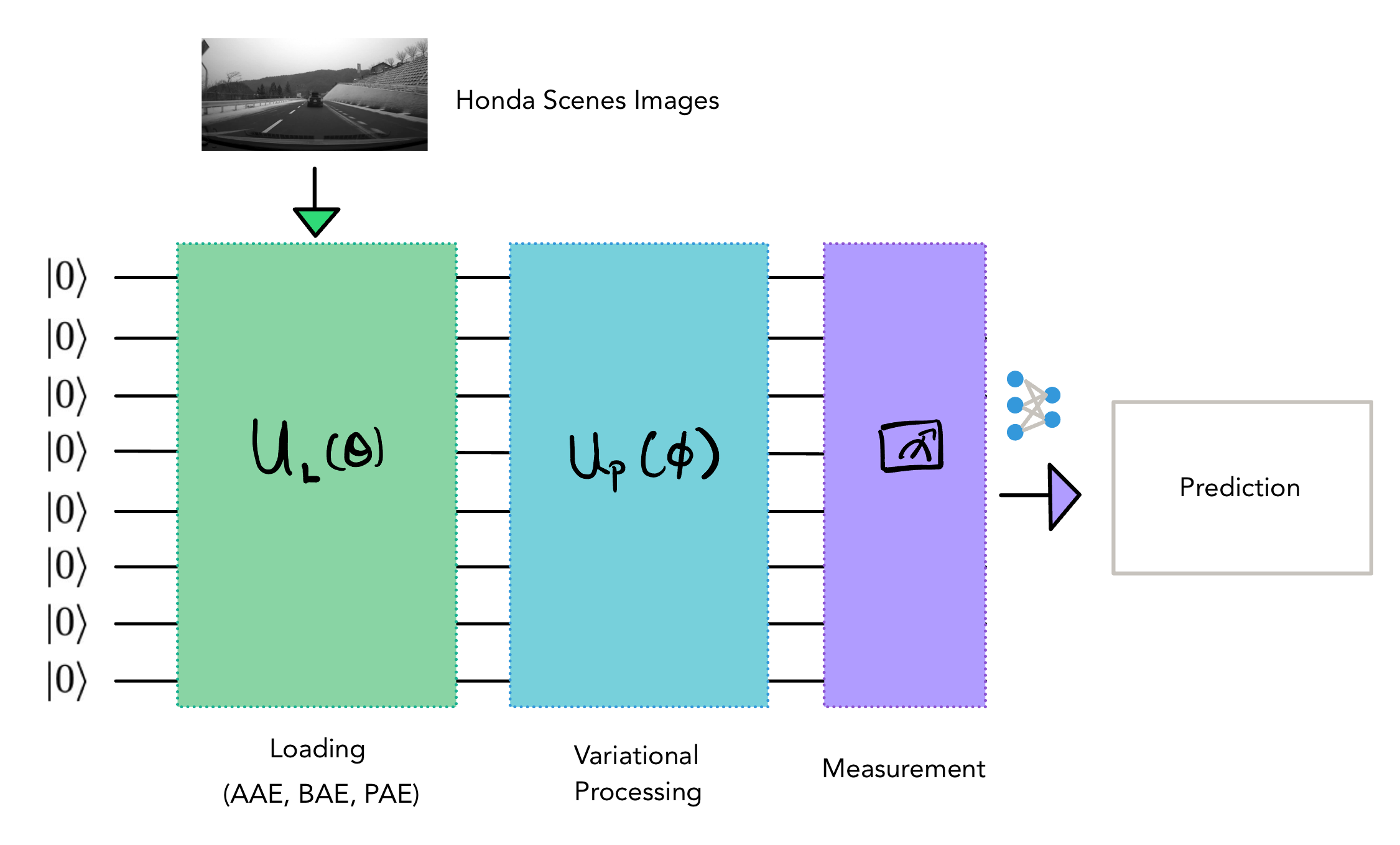}
  \caption{Overview of variational classification circuit. In all of our experiments, they consist of three components: loading (green), processing (blue) and measurement (purple). The measurement outcomes are then fed into a softmax layer to yield class prediction probabilities. The circuit training is generally done in two stages. For the amplitude encoding methods, the loading circuits are learned by minimizing the KL divergence between the prepared input state and the target state. Once the loading is converged, those parameters are fixed while the processing circuit parameters and classical weights from the measurements are trained to minimize the cross entropy loss.}
  \label{fig:variational_circuit}
\end{figure}

In this section, we provide an overview of the methodology for the quantum classification circuits we run. In general, these circuits consist of loading, processing, and measurement components. We begin by describing our choice of images, which come from the Honda Scenes Dataset~\cite{honda-data-set}. Next, we explain different methods of loading the classical image data into the quantum circuit. Finally, we describe the variational classification layers that are common amongst all of the circuits as well as the way measurements will be processed to produce a classification label. An overview is shown in Figure~\ref{fig:variational_circuit}.

\subsection{Honda Scenes Dataset}

In this paper, we focus on the Honda Scenes dataset~\cite{honda-data-set,DBLP:journals/corr/abs-1905-12708}
The Honda Scenes is an extensive, labeled collection designed for dynamic scene classification. It encompasses 80 hours of varied, high-quality driving footage from the San Francisco Bay area. This dataset features time-based annotations detailing various aspects such as road locations, surrounding environment, weather conditions, and the state of the road surface. We will focus primarily on the \textit{Road Surface} classification problem.

\subsection{Data loading}
Every quantum machine learning algorithm must decide how data is represented in the quantum computation. Here we detail the main approaches implemented in this work. Our input data is images. In the case of approximate and block amplitude encoding, the data must be pre-processed to satisfy the normalization requirement of wave functions. In contrast, this type of preprocessing is not necessary for the piece-wise angle encoding.

\subsubsection{Approximate Amplitude Encoding (AAE)}
Given a classical vector $\vec{x}$ of dimension $N$, one can represent it with a state vector on $\lceil\log_{2} N \rceil$ qubits whose entries are rescaled to satisfy the normalization constraint, e.g. $\tilde{x}_{i} = \frac{x_{i}}{|\vec{x}|}$. For $2048 \times 1024$ images, this would require $21$ qubits. When using a fewer number of qubits, we will first downsample the image.

To perform exact amplitude encoding requires exponential number of gates~\cite{Long_2001, Plesch_2011}. However, to perform the actual loading on a quantum device, a shallower circuit is preferred. Thus, we turn to approximate amplitude encoding~\cite{Nakaji_2022}. Our approach differs slightly in that our variational circuit uses a hierarchical ansatz as in~\cite{gharibyan2023hierarchical,honda1}. The intuition behind the hierarchical ansatz is to utilize the structure of the encoding from the image data to the wave function elements. With the typical bitstring mapping, qubits that represent more significant bits play an outsized role in the entanglement structure of the final wave function. The hierarchical ansatz thus does the variational training in steps, first incorporating the most significant qubits before gradually including more.

To find the variational parameters, we use simulators to minimize the KL divergence between the prepared and target amplitudes. We extend the adjoint~\cite{jones2020efficient} techniques that are available in the Pennylane library~\cite{bergholm2022pennylaneautomaticdifferentiationhybrid} and use the Adam optimizer~\cite{DBLP:journals/corr/KingmaB14} for our gradient descent. This of course relies on having access to the state vector, so the training must be done classically. Ref~\cite{Nakaji_2022} suggests training on the maximum mean discrepancy (MMD) cost, similarly to the quantum circuit Born machine (QCBM)~\cite{Liu_2018} where the goal is to encode a probability distribution in the measurement outcomes for the variataional state. With QCBMs, the measurements do not contain the relative phase information, so~\cite{Nakaji_2022} improves on this by also including the Hadamard transformed basis in the MMD loss. We leave the investigation of the sample complexity required for convergence using experimental outcomes to future work. However, it should be noted that in all cases, once the variational parameters for loading the data are learned (classically or quantumly), they can be reused in all further experiments.

\subsubsection{Block Amplitude Encoding (BAE)}
\begin{figure}
    \centering
    \includegraphics[width=0.3\textwidth]{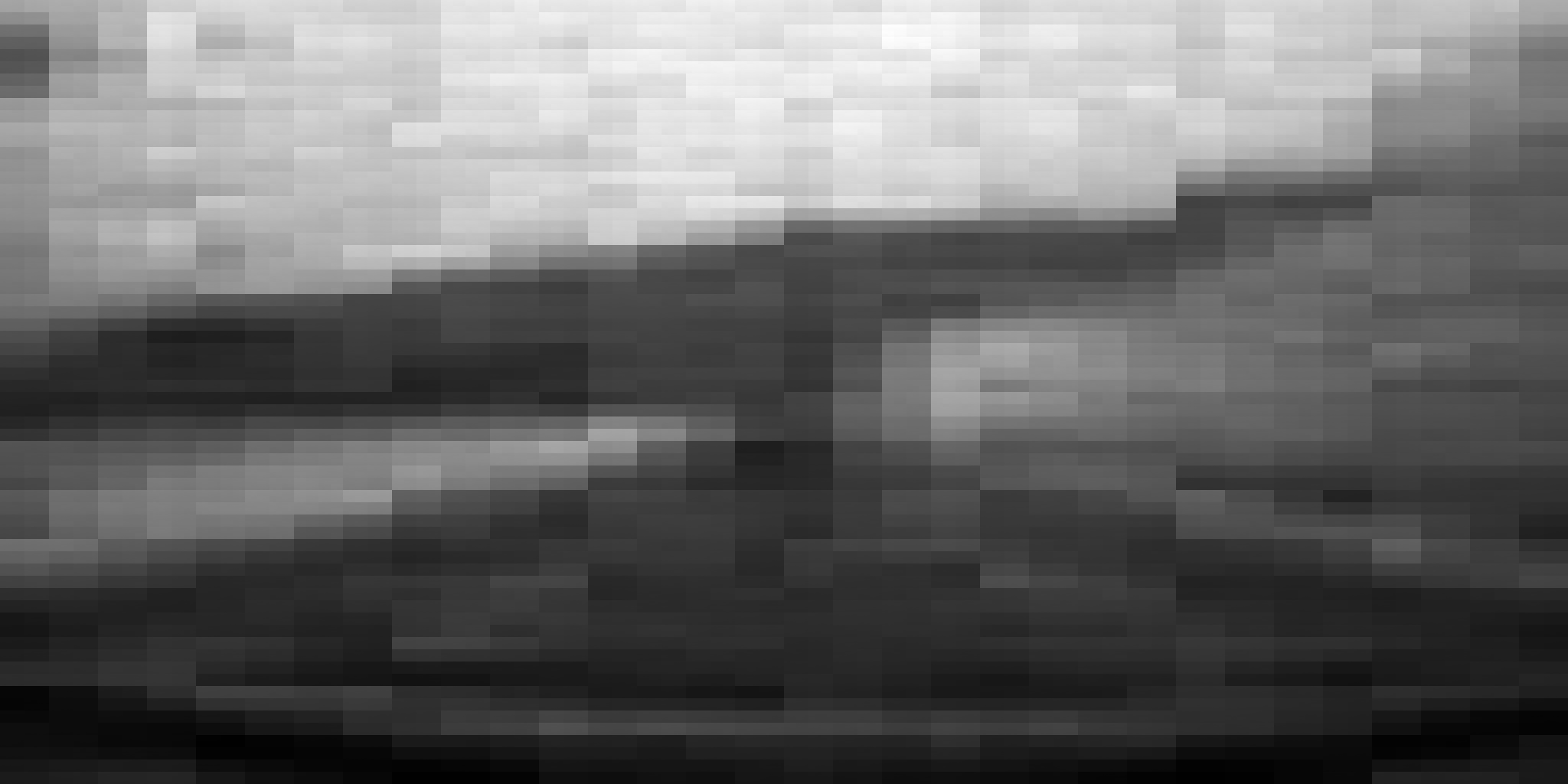}
    \includegraphics[width=0.3\textwidth]{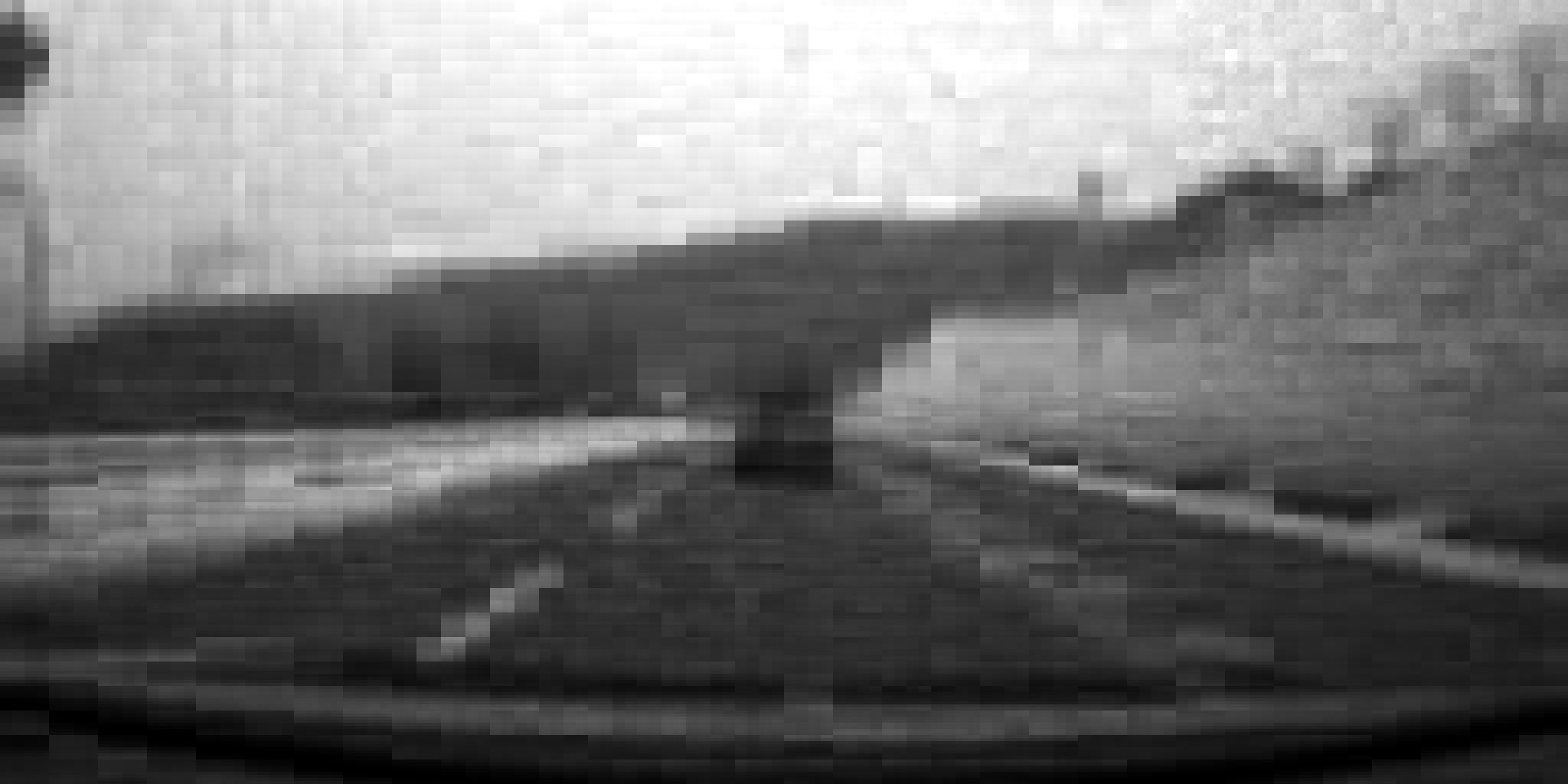}
    \includegraphics[width=0.3\textwidth]{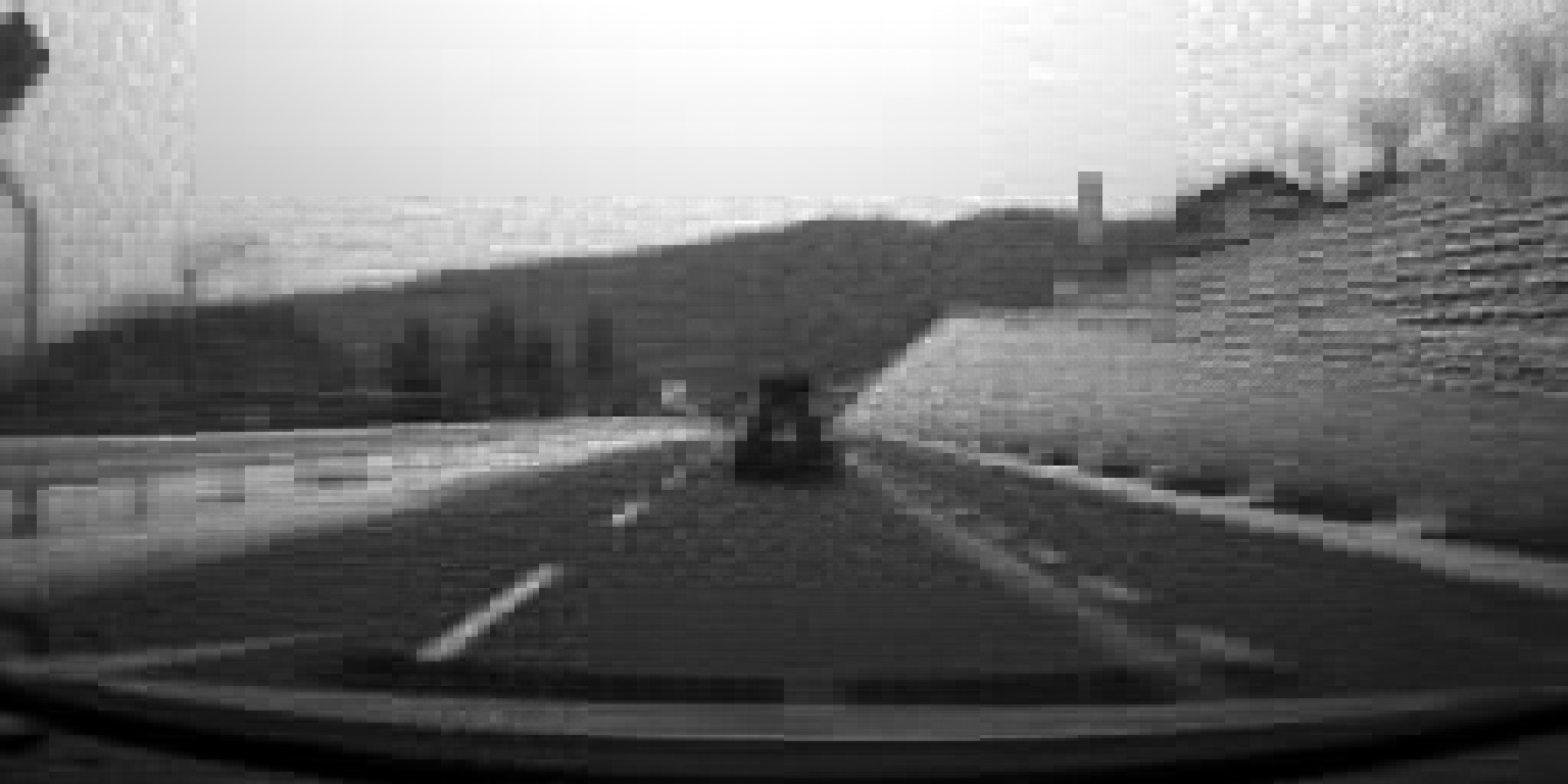}
    \includegraphics[width=0.3\textwidth]{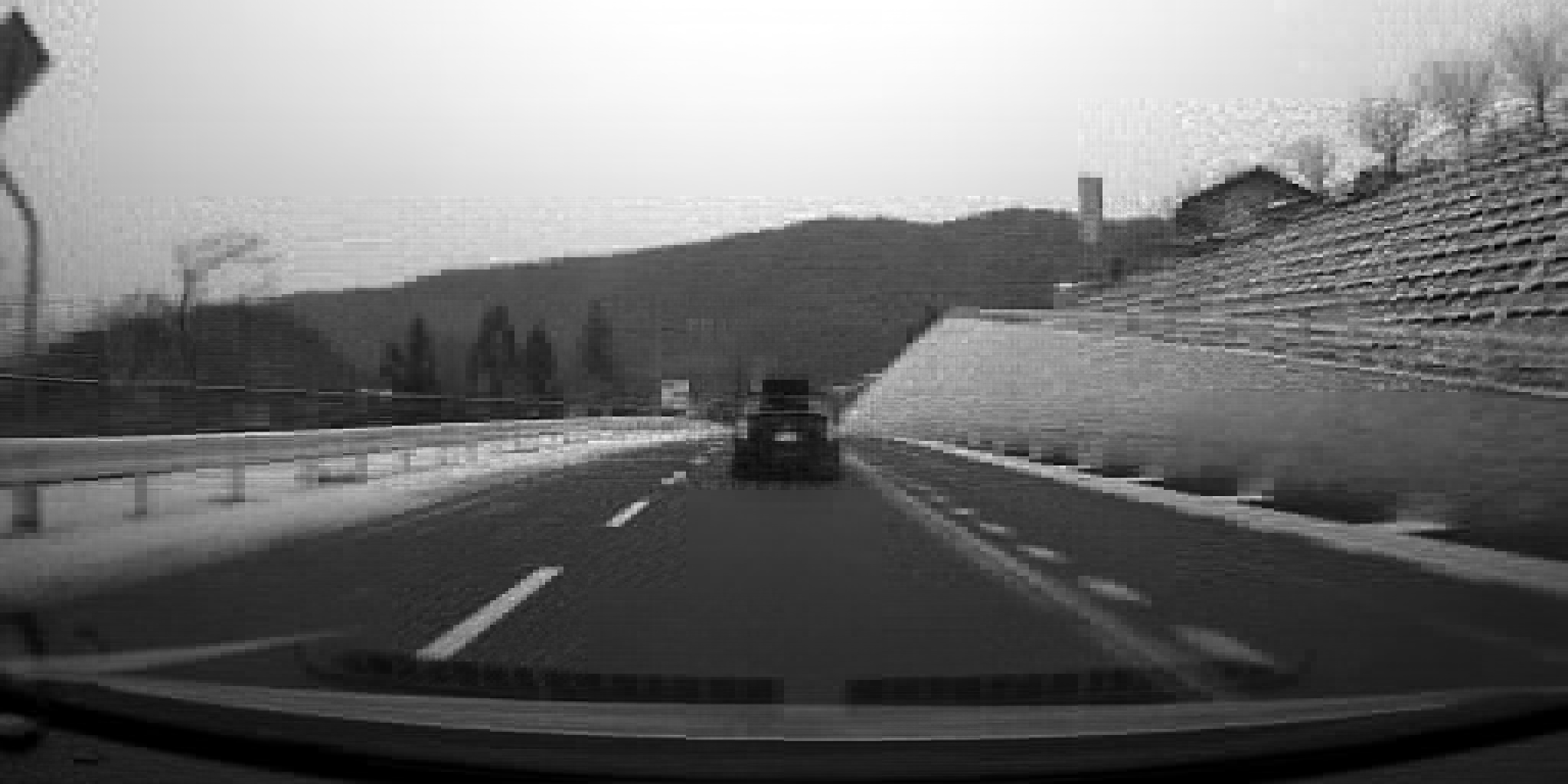}
    \includegraphics[width=0.3\textwidth]{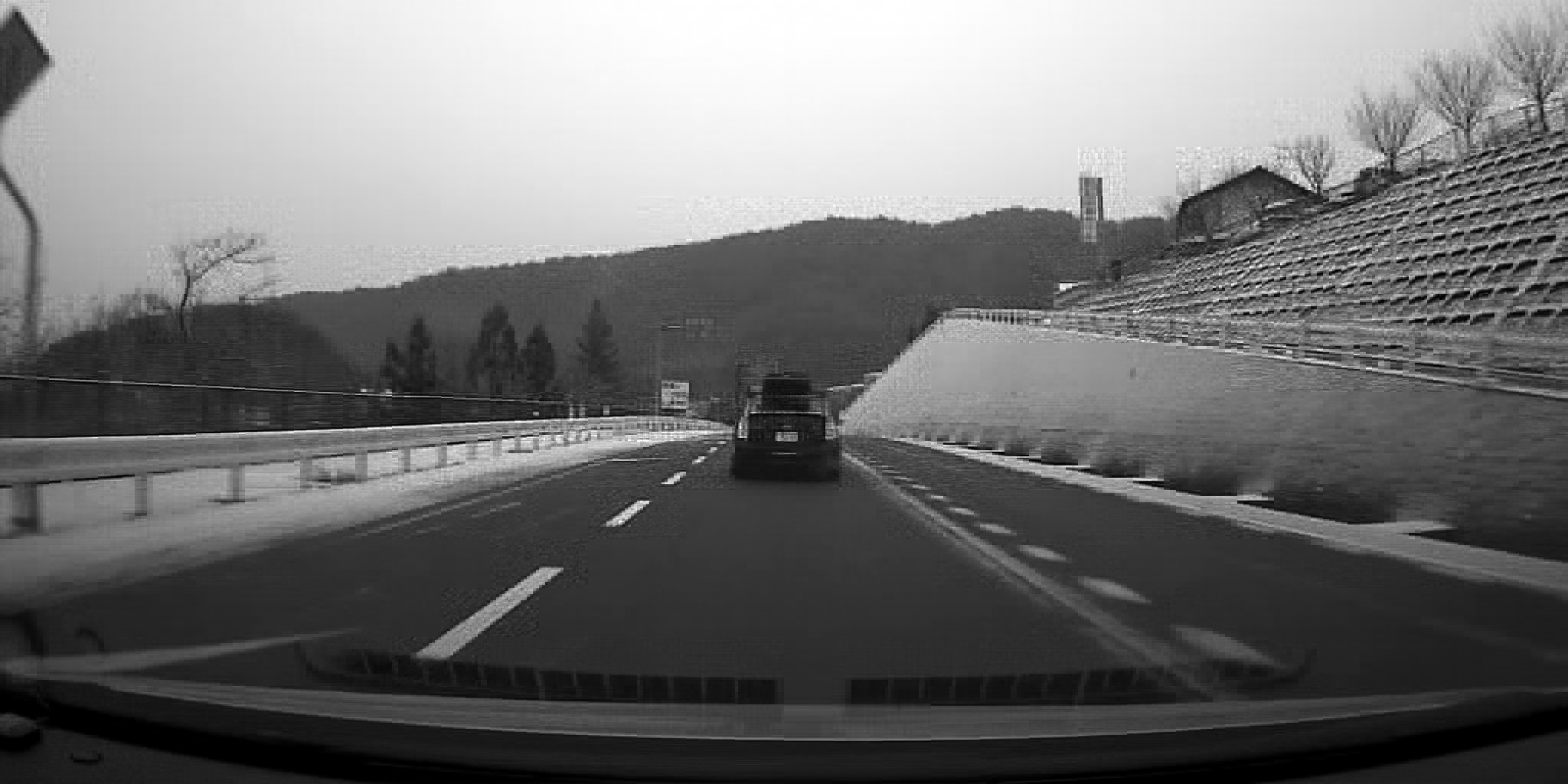}
    \includegraphics[width=0.3\textwidth]{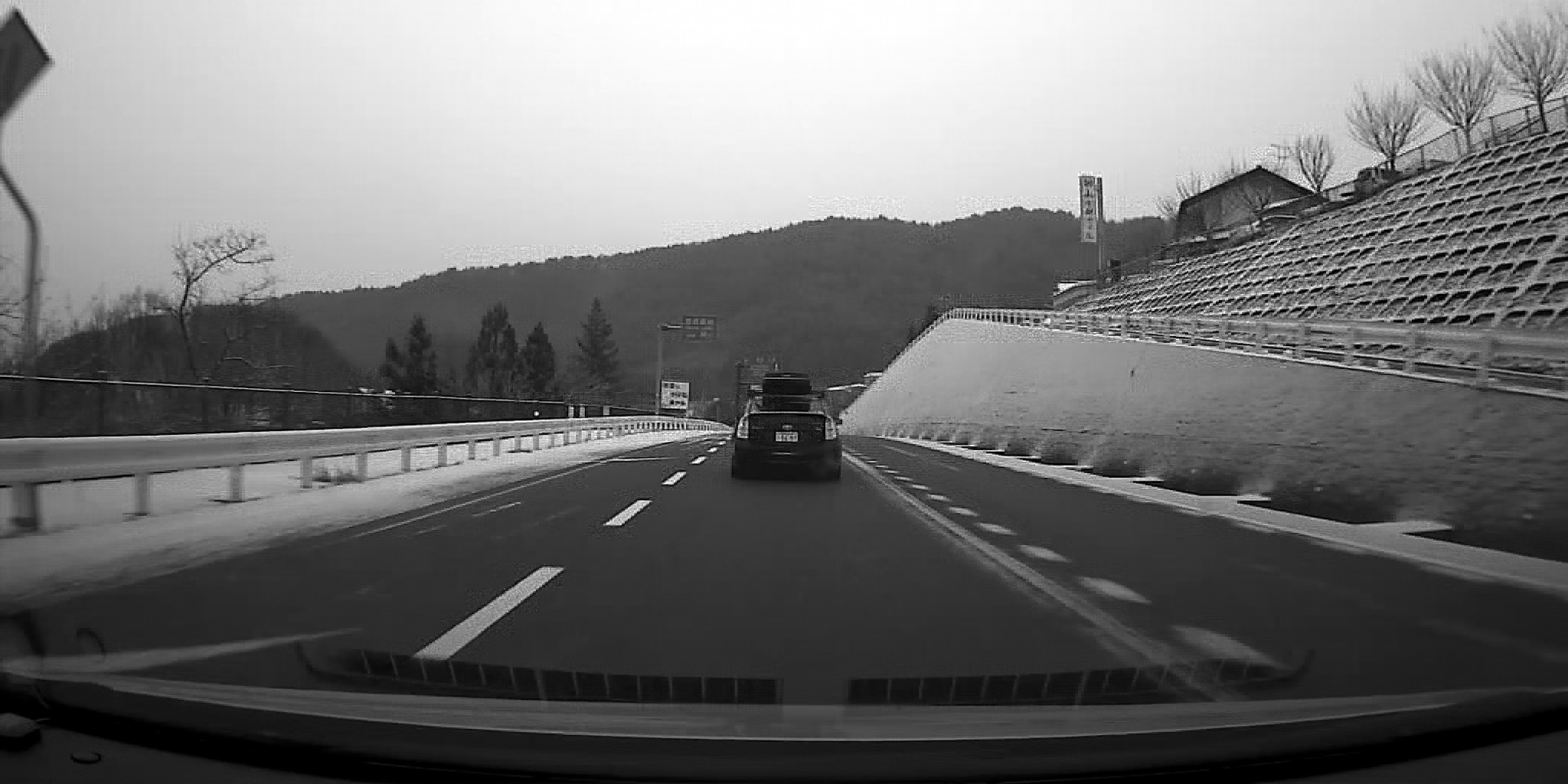}
    \caption{From~\cite{honda1}, the same image from HSD BAE loaded with varying block size. From the top left to bottom right images: 1 block, 21 qubit; 8 blocks, $18 \times 8=144$ qubits; 32 blocks, $16\times32 = 512$ qubits; 128 blocks, $14\times 128 = 1792$ qubits; 512 blocks, $12 \times 512 = 6144$ qubits; 2048 blocks, $10 \times 2048 = 20K$ qubits }
    \label{fig:q-images-bvals}
\end{figure}

Block amplitude encoding (BAE) refers to first partitioning the image into several blocks and performing amplitude encoding on each of them. The global wave function is then a product of these individual blocks. This approach provides a tradeoff in that increasing block number requires larger number of qubits though generally provides a more accurate encoding, since the size of individual blocks is smaller. Another benefit for the near term is that these blocks can be simulated independently and generally loaded in parallel, demanding less depth in quantum hardware. Our notation for the sizes will be $q \times b$ where $q$ is the number of qubits per block and $b$ is the number of blocks. See Figure~\ref{fig:q-images-bvals} for an illustration of loaded images from~\cite{honda1}. When using blocks, the amplitudes corresponding to different parts of the image will no longer be entangled.

\subsubsection{Piecewise Angle Encoding (PAE)}
Angle encoding offers a different way of loading quantum data that specifies the parameters of the varational gates in the loading circuit, as opposed to the amplitudes of the target wave function. The number of qubits here can be variable since the size of the wave function is not tied to the dimension of the input vector. 

In a fashion inspired by re-uploading~\cite{PerezSalinas2020datareuploading}, we will spread the loading parts of the circuit interspersed with the classifier parts of the circuit. This effectively allows for the rotational encoding of the data to not be simply additive across different parts of the image. We experimented with doing all of the encoding all upfront, but it provided worse results.
The number of uploading layers will be determined by the size of the input data and the number of qubits. Each uploading layer consists of 3 encoding rotations per qubit (RY-RZ-RY). This directly lower bounds the number of variational processing layers as well.

\subsection{Data Processing and Classification}

Separately from the encoding, we pick a hardware efficient ansatz (HEA)~\cite{Kandala_2017} for the variational circuit that does the processing. The choice of two-qubit gates is motivated by hardware considerations, and the connectivity likewise is chosen with the hardware layout in mind. A single layer generally includes single qubit rotation gates with two qubit gates between all neighbors. At the end of the circuit, we measure the expectation values of a few qubits. These expectation values are then fed into a classical fully connected layer and used to output class probabilities. In this hybrid setup, we use adjoint methods to compute the gradients with respect to the cross entropy loss.

In the case of the block encoding, the circuit is chosen to be a typical HEA ansatz within the block, though one could in principle add entangling gates between them. One qubit per block is then measured before feeding the expectation value into the final classical layer.

\section{Simulation of Quantum Classifiers}\label{sec:simulation}
\begin{figure}
    \centering
    \includegraphics[width=0.9\textwidth]{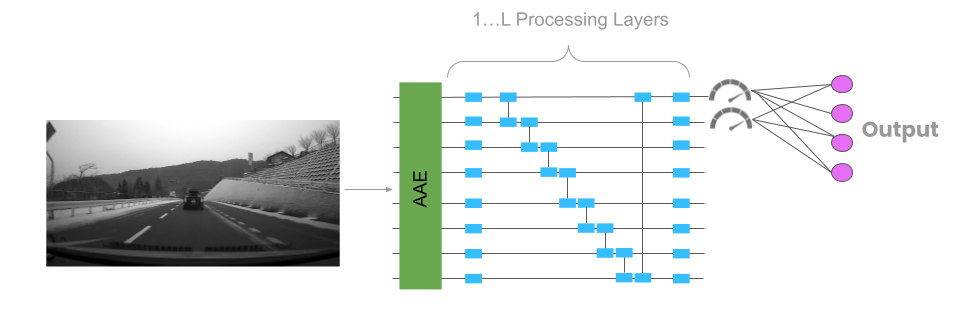}
    \includegraphics[width=0.86\textwidth]{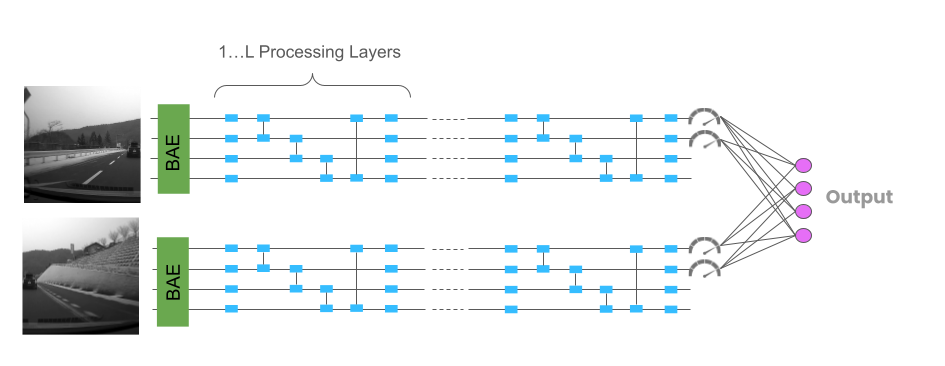}
    \includegraphics[width=0.9\textwidth]{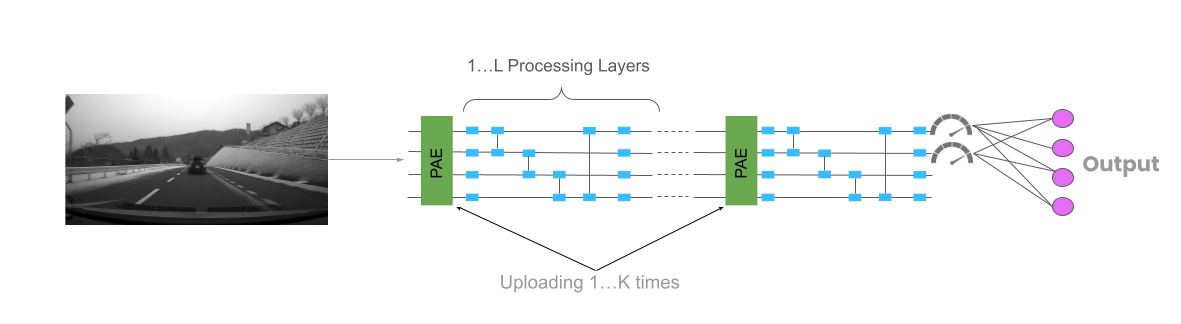}
    \caption{Classifier circuits for different data loading schemes. Top: AAE loading is done across all the qubits first and then the variational layers in blue are learned. Middle: With BAE, the image is partitioned into segments. The corresponding sections of the image are loaded and processed separately. However, their measurement outcomes are combined in the final softmax. Bottom: In the PAE scheme, the data loading layers are interspersed with the processing layers. The number of processing layers is lower bounded by the number of loading layers that are required based on the input size and number of qubits.}\label{fig:loading-classification-circuit}
\end{figure}

In this section we investigate the role of different data loading techniques on a quantum classifier based on variational quantum circuits, as shown in Figure~\ref{fig:loading-classification-circuit}. We focus on a multi-class dataset of 2,500 images with four classes: clear day, clear evening, snowy day, and snowy evening. In all of our circuit classifiers, there is no classical feature extraction before the quantum processing. Rather than encoding the prediction in the measurement outcome of a single qubit, we apply a small linear layer to expectation values of a few qubits followed by a softmax to output class probabilities. For comparison, we also implement a variety of classical neural networks. The results are presented in Table~\ref{tab:sorted_comparison}.

\begin{table}[h!]
    \centering
    \begin{tabular}{|l|c|r|l|c|c|}
        \hline
        \textbf{Type} & \textbf{Model} & \textbf{Params} & \textbf{Architecture} & \textbf{Acc} & \textbf{FLOPs/QUOPs}\\ \hline
        \textbf{Classical} & \textbf{CNN} & \textbf{1994} & \textbf{layers of CNN, maxpool, FC} & \textbf{0.97} & 10M \\ \hline
        Classical & FCN & 8196 & input-fc-fc-output & 0.93  & 10K \\ \hline
        Classical & CNN & 297 & layers of CNN, maxpool, FC & 0.92 & 2M \\ \hline
        \textbf{Quantum} & \textbf{BAE} & \textbf{3652} & \textbf{14$\times$16=224 qubits, 4$\times$4 blocks} & \textbf{0.94} & 10K \\ \hline
        Quantum & BAE & 2052 & 14$\times$8=112 qubits, 2$\times$4 blocks & 0.93 & 5K\\ \hline
        Quantum & BAE & 884 & 12$\times$4=48 qubits, 2$\times$2 blocks & 0.91 & 2K\\ \hline
        Quantum & PAE & 372 & 20 qubits, 16$\times$24 input & 0.88 & 1K\\ \hline
        Quantum & PAE & 332 & 16 qubits, 16$\times$24 input & 0.87 & 1K\\ \hline
        Quantum & BAE & 308 & 12$\times$4=48 qubits, 2$\times$2 blocks & 0.82 & 1K \\ \hline
        Quantum & AAE & 354 & 19 qubits, 6 layers & 0.81 & 1K \\ \hline
        Quantum & PAE & 876 & 16 qubits, 16$\times$24 input & 0.76 & 2K \\ \hline
        Quantum & AAE & 246 & 13 qubits, 6 layers & 0.73 & 1K\\ \hline
    \end{tabular}
    \caption{Comparison of Classical and Quantum Models on the four class scene classification task. On the classical side, we try state of the art convolutional neural networks (CNNs) as well as more simple fully-connected (FC) layers. While the largest CNN has fewer parameters than the largest quantum model, the architecture is significantly more complex. In the last column we report approximate FLOPs with Quantum OPs~\cite{preskill-quop} to compare model complexity.}
    \label{tab:sorted_comparison}
\end{table}

We find that the best performance in the quantum models is achieved with the BAE loading method. Our largest model involved 16 blocks each consisting of 14 qubits and 6 variational processing layers yielding an accuracy of $0.94$. Notably, we find that this is slightly worse than a convolutional neural network with fewer parameters but better than a vanilla fully-connected neural network with more weights.

The next best performing quantum model was the piecewise angle encoding. To limit the number of uploading layers needed in the circuit, the image is significantly downsampled. Interestingly, the larger model with more processing layers in between the uploading layers performs worse. Finally, the AAE method yielded the worst performance.

We note that our findings of performance for the different data loading methods are likely not definitive. For example, both BAE and AAE are approximate by nature, so it is possible that more accurate variational circuits for the loading could improve performance. Similarly, we have fixed the processing layers to be based on the HEA ansatz, whereas different architectures may better suit different loading methods.

\section{Deploying Classifiers on Quantum Hardware}\label{sec:quantum-expts}

In this section we present the results of running classifier models described above on Quantinuum and IBM devices. The models are trained using simulators and subsequently deployed on the hardware for inference runs. Our choice of classification problems and models were motivated largely by hardware considerations. For example, the variational circuits we choose respect the native gate set and connectivity to be hardware friendly. To fit in relatively short depth circuits, we simplify the classification to be binary rather than multi-class, distinguishing between \texttt{clear} vs \texttt{snow} road conditions.

Our experiments vary in their method of loading the data as well as the hardware, and thus the structure of variational circuits. For each of the runs, the full pipeline includes first training the quantum circuits on classical simulators (CPU or GPU) on the training set. The variational parameters are then saved and deployed on quantum hardware. We track the performance of the hardware itself by comparing measurement outcomes to values produced by the simulator. The Quantinuum H1 and H2 systems have 20 and 56 qubits, respectively, allowing us to run amplitude encoding and piecewise angle encoding experiments. The IBM Brisbane and Fez devices, with 127 and 156 qubits, allow us to additionally explore the block amplitude encoding method. Our largest circuits utilize thousands of two-qubit gates.
\begin{figure}
  \centering
  \includegraphics[width=.4\textwidth]{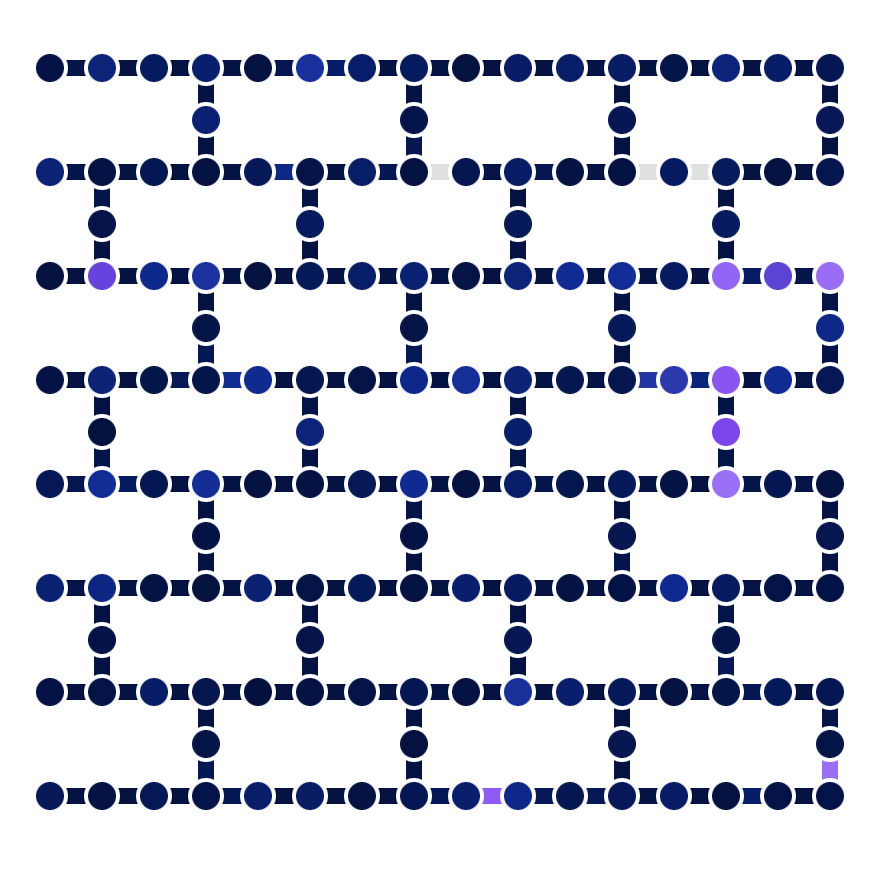}

  \caption{Qubit connectivity for the 156 qubit IBM Fez device. For block amplitude encoding, we perform the loading and quantum variational layers in parallel. We take advantage of the heavy hex layout by using the ring connectivity for blocks of 12 qubits. Colored circles denote qubits with readout errors (darker is lower) and the edge color denotes the median CZ gate error. Accessed from IBM website August 16, 2024. }
\end{figure}

\subsection{Quantinuum experiments}
Quantinuum System Model H1 is a 20-qubit universal quantum computer based on trapped-ions. It has all-to-all connectivity, at the time of the testing a reported median 1-qubit fidelity of $99.996\%$ and median 2-qubit fidelity of $99.9\%$. The H2 is a 56-qubit device with similarly impressive 1 and 2-qubit fidelities at $99.997\%$ and $99.87\%$~\cite{Moses_2023}. With limited access to both devices, we run inference on just a handful of images after training the classifier with simulators.

\subsubsection{Quantinuum Model H1}
The circuits we run involve two of the data loading methods, AAE and PAE, with variational classification layers. The measurement is done on the first two qubit registers and we feed their expectation values into a linear classifier. In evaluating the performance of the hardware, we track the two Z-observable expectations of the two measured qubits, and the output probability of the inference. The output probabilities are obtained after doing a linear map on the expectation values to obtain logits and then applying softmax on top of these 2 numbers to obtain final output probabilities, as in a vanilla neural network output layer.

For both AAE and PAE experiments, the images are downsampled depending on the qubit numbers. We have varying sizes with $N=10, 14$ and $20$. For AAE, we downsample the image to be $2^{N/2}$ x $2^{N/2}$ so it fits in the $N$ qubit wave function and then add 3 variational layers as a classifier. For PAE, we downsample the image to a smaller size such that it can be encoded in the rotation angles $<10$ uploading layers, after which we have another 3 layers for classification.

Due to very limited access to the hardware, for each of the models, we pick a single \texttt{clear} image and run inference on the H1 machine. Not only did we get back the correct class with $>90\%$ probability each time, but we also noticed a very little difference between bitstring probabilities compared to simulator runs. The two qubit gate counts ranged from $\sim 60$ up to $420$. Table~\ref{tab:quantinuum-h1-runs} shows the results of these 6 experiments.

\begin{table}[]
  \resizebox{\textwidth}{!}{%
    \begin{tabular}{|c|c|c|c|c|c|c|} \hline
      \textbf{Model} & \textbf{\# Qubits} & \textbf{\# 1Q gates} & \textbf{\# 2Q gates} & \textbf{Model Accuracy} & \textbf{$P_{sim}$(\texttt{clear})} & \textbf{$P_{quantum}$(\texttt{clear})} \\ \hline
      AAE            & 10                & 146                 & 66                  & 0.88                    & 0.99                              & 0.998                                              \\ \hline
      AAE            & 14                & 226                 & 112                 & 0.88                    & 0.92                              & 0.90                                               \\ \hline
      AAE            & 20                & 395                 & 205                 & 0.88                    & 0.94                              & 0.97                                               \\ \hline
      PAE            & 10                & 500                 & 120                 & 0.997                   & 0.996                             & 0.998                                              \\ \hline
      PAE            & 14                & 770                 & 180                 & 0.997                   & 0.99                              & 0.99                                               \\ \hline
      PAE            & 20                & 1900                & 420                 & 0.997                   & 0.99                              & 0.97                                               \\ \hline
    \end{tabular}%
  }
  \caption{Results of the 6 inference runs on Quantinuum H1 device. Model accuracy is the score of the model on a test dataset after training. $P(\texttt{clear})$ is the probability of outputting the correct class of a particular \texttt{clear} image using the learned circuit in simulator and on hardware. This probability is based on the softmax layer applied to expectation values of two qubits.}
  \label{tab:quantinuum-h1-runs}
\end{table}

\subsubsection{Quantinuum Model H2}

\begin{figure}
    \centering
    \includegraphics[width=0.85\textwidth]{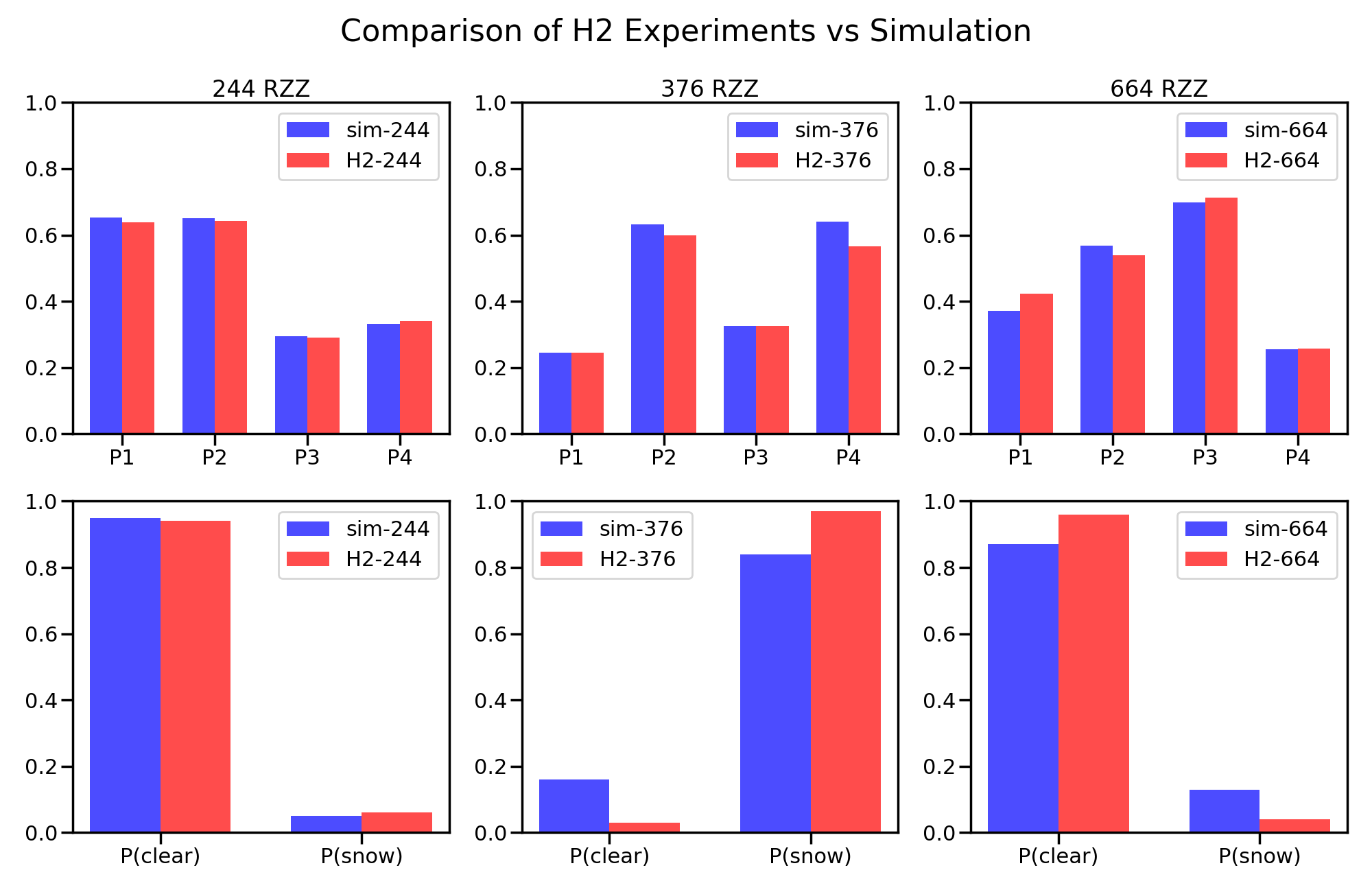}
    \caption{Performance of H2 classifier circuits of different depths with BAE loading. We track the probabilities of individual expectation values for 4 qubits, the most significant qubit of each block as illustrated in Fig.~\ref{fig:loading-classification-circuit}. In the second row, we plot the predicted class probabilities for the image based on the expectation values. The respective shot numbers for the 244 RZZ, 376 RZZ and 664 RZZ circuits were 400, 200, and 300. In all cases, the classifier predicts the correct class.}
    \label{fig:h2-runs}
\end{figure}

The larger Model H2 device, with 56 qubits, allowed us to explore the BAE data loading method with four blocks of 14 qubits each.
We ran three circuit architectures ranging from 244 to 664 two-qubit gates each running inference on a different image. The performance was quite promising as seen in Figure~\ref{fig:h2-runs}. For the classifier, the expectation value of the most significant qubit for each of the blocks is measured. Equivalently, we track the probability of measuring the qubit in the 0 state, denoted $P_i$. In addition to the comparison of simulated and obtained expectation values, we also plot the classifier probabilities. The sum of absolute deviations, which we denote as $L_{1} = \sum_{i} |P_{i}^{{sim}} - P_{i}^{{expt}}|$, across the three circuit sizes are $0.035, 0.108 $ and $0.074$. These are not necessarily directly comparable for two reasons but they give an idea of the performance. First, the number of shots was different, with 400, 200 and 300 shots from smallest to largest. Thus, it's possible that the $L_1$ for the medium-sized circuit may be the largest due to having lower shot number. Second, each of the circuits was trained separately, so the underlying $P_i$ are different for each circuit.

\subsection{IBM Experiments}
We present a series of experiments carried out on the Brisbane and Fez devices.
IBM Brisbane is a 127 qubit universal quantum computer that has median 99.9\% 1-qubit gate fidelity and median 99.2\% 2-qubit gate fidelity based on the Eagle chip. Fez is a next generation chip with up to 156 qubits, median 99.9\% 1-qubit gate fidelity and median 99.7\% 2-qubit gate fidelity. Due to the relative affordability, we were able to run more extensive experiments. As before, we perform the simulation on an ansatz which is tailored to the devices' connectivity and native gate sets. The circuits are furthur optimized for error mitigation at the quantum control level with the Q-CTRL software. However, slightly differently from the last section, we will run inference on 100 images consisting of 50 from each class. Since the qubit counts are higher, it allowed more flexibility in experimenting with different block sizes for the BAE loading.

\subsubsection{IBM Brisbane}
With Brisbane, we deployed circuits BAE and PAE data loading methods.
We use CNOT gates as entanglers, ring connectivity to match the sparse grid-like structure of IBM Brisbane and use alternating brickwork entangling layers. The latter makes sure we achieve reasonable entanglement without going too deep. We carry out 2 experiments with PAE, utilizing 12 and 16 qubits. The model accuracy scores for these are 96\% and 97\% respectively. For BAE, we tried a model with 6 blocks of 12 qubits each --- total of 72 qubits --- making this the largest model in terms of number of qubits we have run. The test score for this model is 99.3\%. Again, for each experiment, we run inference on 100 images and report the results in Table~\ref{tab:brisbane-runs-classifier}.
As we see the 12 and 16 qubit experiments almost perfectly match the simulator results, whereas there is a noticeable degradation for the 72 qubit circuit where accuracy drops from 0.99 to 0.7.

\begin{table}[]
\resizebox{\textwidth}{!}{%
\begin{tabular}{|c|c|c|c|c|c|c|c|}
\hline
\textbf{Model Type} & \textbf{\# Qubits} & \textbf{\# 1q gates} & \textbf{\# 2q gates} & \textbf{Depth} & \textbf{Model Acc} & \textbf{Sim Acc} & \textbf{Brisbane Acc} \\ \hline
PAE                 & 12                 & 300                  & 36                   & 32             & 0.94             & 0.97             & 0.98             \\ \hline
PAE                 & 16                 & 600                  & 75                   & 52             & 0.97             & 0.97             & 0.95             \\ \hline
BAE                 & 12 $\times$ 6                 & 2200                 & 276                  & 50             & 0.993            & 0.99             & 0.7              \\ \hline
\end{tabular}%
}
\caption{Results of the 3 experimental runs on IBM Brisbane device. Model accuracy is the accuracy trained model on the test dataset using simulators. For the device runs, we have 100 images, 50 from each class. IBM accuracy is based on the classification outcomes of those 100 images based on hardware. Simulation accuracy is based on that same set of images. }
\label{tab:brisbane-runs-classifier}
\end{table}

\subsubsection{IBM Fez}

On IBM Fez, we push the device limits even further by executing our BAE experiments up to circuits with thousands of CX gates. Similarly to the Brisbane runs, we use 72 qubits. In simulators, all the models were able to succeed on the binary classification task with 99\% accuracy. There also appears to be a bias in the model for getting the accuracy of the \texttt{clear} class over the \texttt{snow} class. Table~\ref{tab:fez_summary} contains the circuit information for various experiments sorted by total two-qubit gate count. We vary both the size of the loading circuit as well as the classification circuit. Since all models have the same simulation accuracy, we can see the general trend that the accuracy declines with increasing circuit depth. Impressively, we see that even with over 3000 two-qubit gates, we manage to correctly classify 68\% of the images. 

\begin{figure}
    \centering
    \includegraphics[width=0.6\linewidth]{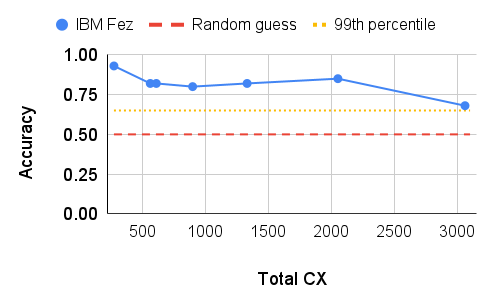}
    \caption{The accuracy of BAE loaded circuits on IBM Fez as a function of circuit depth. A random guess would lead to 50\% accuracy on the balanced dataset. Using Monte Carlo sampling, we find that the 99th percentile of outcomes for random guessing would lead to an accuracy of 65\%, lower than the accuracy even from our largest circuit with over 3000 two-qubit gates.}
    \label{fig:fez_accuracy}
\end{figure}

A random classifier would achieve an accuracy of 50\%, so one may ask how meaningful our results are relative to having no model. Using simple Monte Carlo with a binomial, we can estimate the range of outcomes. We find that the 99th percentile of accuracy with a circuit completely overwhelmed by noise would yield an accuracy of 0.65, indicating that we are still getting a meaningful signal out of our largest circuit. See Fig.~\ref{fig:fez_accuracy}.

\begin{table}
\resizebox{\textwidth}{!}{%
\begin{tabular}{|c|c|c|c|c|c|c|c|c|c|}
\hline
\textbf{Total CX} & \textbf{Total Gates}  & \textbf{Depth} & \textbf{\# Layers} & \textbf{\# Params} & \textbf{\texttt{clear} Acc} & \textbf{\texttt{snow} Acc} & \textbf{Fez Acc} & \textbf{Sim Acc}\\
\hline
$132 + 144 = 276$ & 780 $\rightarrow$ 2022 & $18\rightarrow 46$ &  2 & 302 & 1 & 0.87& 0.93 & 0.99\\ \hline
$132 + 432 = 564$ & 1644 $\rightarrow$ 4038 & $34 \rightarrow 78 $ & 6 & 878 & 1 &  0.63&0.82& 0.99 \\ \hline
$468 + 144 = 612$ & 1476 $\rightarrow$ 4249 & $36 \rightarrow 96$ & 2 & 302 & 1 & 0.63& 0.82 & 0.99\\ \hline
$468 + 432 = 900$ & 2340 $\rightarrow$ 6265 & $52 \rightarrow 128$ & 6 & 878 & 0.97 & 0.63& 0.8 & 0.99\\ \hline
$468 + 864 = 1332$ & 1908 $\rightarrow$ 8363 & $46 \rightarrow 191$ & 6 & 878 & 0.77 & 0.87& 0.82 & 0.99\\ \hline
$468 + 1584 = 2052$ & 2628 $\rightarrow$ 12603 & $61 \rightarrow 280$  &  11 & 1598 & 0.9 & 0.8 & 0.85 & 0.99\\ \hline
$468 + 2592 = 3060$ & 3636 $\rightarrow$ 18513 &  $82\rightarrow 407$ & 18 & 2606 & 0.7 & 0.67 & 0.68 & 0.99\\ \hline
\end{tabular}
}
\caption{Summary of IBM Fez experiments. Total CX count is comprised of the loading circuit and the classifier circuit count. For total gates and depth, arrows indicate the change from simulated circuits to transpiled circuits on hardware. The numbers of layers and parameters refer to the classification layers. All circuits were run with models that performed very well on the test set, so we can effectively see the trend of performance as the number of gates increases. The accuracy if visualized in Fig.~\ref{fig:fez_accuracy}.}
\label{tab:fez_summary}
\end{table}

\subsection{Summary of Quantum Experiments}

These experiments constitute the largest image classification performed on available quantum hardware, both in qubit count (up to 72) as well as circuit depth in the thousands. We reiterate that all of the training was done classically, so there are not claims of a quantum advantage here. However, it is noteworthy that the training could be performed with quantum experiments of this scale. Similarly to the simulation only results, we find that the BAE data loading method performs the best followed by PAE and then AAE.

Our results highlight the robust capabilities of both Quantinuum and IBM devices. Both the H1 and H2 devices had remarkable performance up to several hundred 2-qubit gates, where the measurement outcomes used for the classifier deviated very little from the simulated values, even with only a few hundred shots.
On the IBM side, we saw a marked improvement in experimental results going from Brisbane to Fez. In our largest BAE circuits, Brisbane achieved 70\% accuracy on 100 test images using a circuit with 240 CNOT gates. In contrast, Fez achieved 68\% accuracy on a circuit with over 3000 CNOT gates. Comparing the Quantinuum and IBM performance is difficult since the cost difference meant we could not run inference on the same number of test images, however, we did appear to have higher fidelity in the regime of $\sim$600 2-qubit gates on Quantinuum devices.

\section{Discussion}\label{sec:discussion}

In this work, we have trained and deployed a number of quantum circuit classifiers for the task of scene classification. We experimented with a variety of data loading schemes. While it is tempting to compare with classical methods, the comparison is not so straightforward. Our aim was not to declare any advantage but rather to explore the performance of different methods of data encoding --- each with their own set of advantages and drawbacks.  In general, it is not clear which representation would be easiest to process in the quantum circuit, though we find the block amplitude encoding generally yielded the best results. BAE loading allows for efficient classical simulation since the individual block sizes are modest. Additionally, since our classifier circuits only had entangling gates within the blocks, one could do these experiments block by block rather than utilizing all of them at once. However, an interesting strategy would be to add variational gates between blocks that are learned with gradient methods using hardware, quickly leaving the realm of classical simulability. Additionally, starting with these classically learned weights may provide a useful parameter initialization.

The other primary goal of this work is to showcase the limits of existing quantum hardware. In the era of NISQ devices, the utility of variational methods is still debated. They are the natural candidate given the lack of fault tolerance and intuition for how to design quantum algorithms. However, with theoretical results on vanishing gradients and existing noise rates, it is easy to be pessimistic about their potential on quantum devices. Instead, we give evidence that perhaps this conclusion is too fast. Our learning takes place classically rather than quantumly, owing to simulations enabled by specialized adjoint gradient techniques deployed on GPUs, and we were pleased to see hardware results in reasonable agreement even at depths of thousands of 2-qubit gates. While no claims can be made about quantum advantage based on these results, these experiments show that quantum hardware is well-established enough to run meaningfully deep circuits.

There are many natural follow-up directions to pursue, including varying the data loading methods, testing different classifier methods, and incorporating noise model simulations into the training. As hardware continues to improve and systems become more accessible, doing at least some of the training on the quantum hardware itself would be an impressive feat. Of course, one may also investigate using different classification methods (e.g. kernel methods as opposed to using observables) and hybrid architectures.

\textit{Author Contributions:} H. G. contributed to the concept, ideation, algorithm design, and paper editing. H. K. handled data preparation, data processing, and the implementation of the classical CNN model. T. S. was responsible for IBM deployment, circuit engineering, and evaluation. P. S. contributed to the concept, ideation, data selection, and paper editing. V. P. S. on algorithm design, quantum circuit implementation, and paper editing. R. H. T. focused on engineering work on CPU/GPU systems and evaluation. Finally, H. T. led the concept and algorithm implementation, served as the engineering lead, and managed the Quantinuum deployment.

\newpage
\appendix

\section{Software and Hardware details}
Below is the tech stack used for the experiments described in this paper:

\textbf{Nvidia H100 GPUs for gpu simulation}. For large circuit simulations of 20+ qubits the gpu simulators and Nvidia's cuQuantum library were used to achieve high performance and shorten the runtimes. 

\textbf{IBM's 127 qubit Brisbane and Heron chips}. These are IBM's latest quantum chips that are publicly available and boast $>99\%$ median 2-qubit gate fidelity as well as $>99.9\%$ median 1-qubit gate fidelity.

\textbf{Quantinuum's H1 and H2 chips.} These are the flagship chips of Quantinuum and have remarkable fidelities: H1 has 20 fully connected qubits with $99.998\%$ 1-qubit gate fidelities and $99.9\%$ 2-qubit gate fidelity and H2 has 56 qubits with similar fidelities. Quantinuum uses trapped ions for its qubits which are in general slower than superconducting qubits but instead have better qualities and connectivity.

\textbf{BlueQubit SDK for simulations and overall orchestration.} We used BlueQubit's SDK and platform for seamless orchestration of all the pieces needed in the experiments presented in this paper.

\bibliographystyle{alpha}
\bibliography{refs_gibbs}

\end{document}